\documentclass[apj]{emulateapj}
\usepackage{apjfonts}
\usepackage{amsbsy}

\def\wp{$w_p(r_p)$}
\def\hmpc{$h^{-1}$Mpc}
\def\hkpc{$h^{-1}$kpc}

\def\mstar{$M_\ast$}

\def\hmsol{$h^{-1}$M$_\odot$}
\def\kms{km\,s$^{-1}$}

\def\cmbfast{{\scriptsize CMBFAST}}

\def\navg{\langle N\rangle_M}

\def\nsat{\langle N_{\rm sat}\rangle_M}
\def\ncen{\langle N_{\rm cen}\rangle_M}
\def\mmin{M_{\rm min}}
\def\mcut{M_{\rm cut}}

\def\asat{\alpha_{\rm sat}}

\def\om{\Omega_m}
\def\omb{\Omega_b}
\def\s8{\sigma_8}
\def\av{\alpha_v}

\def\plin{P_{\rm lin}(k)}

\def\lcdm{$\Lambda$CDM}

\def\x2{$\chi^2$}
\def\hmsol{$h^{-1}\,$M$_\odot$}

\def\NNm1{\langle N(N-1) \rangle}

\def\fsat{f_{\rm sat}}
\def\slogm{\sigma_{{\rm log}M}}
\def\m_star{M_\ast}

\def\plin{P_{\rm lin}(k)}

\def\xirp{\xi(r_p,r_\pi)}

\def\navgi{\langle N\rangle_M^i}
\def\nceni{\langle N_{\mbox{\scriptsize cen}}\rangle_M^i}
\def\ncenip{\langle N_{\mbox{\scriptsize cen}}\rangle_M^{i+1}}

\def\mmini{M_{\rm min}^i}
\def\mcuti{M_{\rm cut}^i}
\def\sigmaMi{\sigma_{\log M}^i}
\def\slogmi{\sigma_{\log M}^i}
\def\nsati{\langle N_{\mbox{\scriptsize sat}}\rangle_M^i}
\def\monei{M_1^i}
\def\mONEi{M_{\rm one}^i}
\def\ngavgi{\bar{n}_g^i}

\def\mbj{M_{b_J}}

\def\mpch{Mpc$^{-1}\,h$}

\def\fsat{f_{\rm sat}}
\def\fsati{f_{\rm sat}^i}

\def\sk{\sigma_k}
\def\sv{\sigma_v}

\def\pz{P_Z(k,\mu)}
\def\mstar{M_\ast}

\begin{document}

\title{On the Luminosity Dependence of the Galaxy Pairwise Velocity Dispersion}

\author{Jeremy L. Tinker\altaffilmark{1},
Peder Norberg\altaffilmark{2},
David H. Weinberg\altaffilmark{3}
\& Michael S. Warren\altaffilmark{4}
}
\altaffiltext{1}{Kavli Institute for Cosmological Physics, University of Chicago}
\altaffiltext{2}{Institute for Astronomy, University of Edinburgh}
\altaffiltext{3}{Department of Astronomy, Ohio State University}
\altaffiltext{4}{Theoretical Astrophysics, Los Alamos National Labs}

\begin{abstract}

We make predictions for the pairwise velocity dispersion (PVD) of
galaxies with models that are constrained to match the projected
correlation function and luminosity function of galaxies in the
Two-Degree Field Galaxy Redshift Survey (2dFGRS). We use these data to
constrain the halo occupation distribution (HOD) of 2dFGRS galaxies,
then calculate the PVD by populating the halos of a large-volume,
high-resolution N-body simulation. We examine the luminosity and scale
dependence of the predicted PVD. At large scales the PVD is flat, but
at $r\sim 2$ \hmpc\ the dispersion rapidly rises as the dominant
source of galaxy pairs transitions from two-halo pairs to pairs within
a single halo. At small and large scales, $r< 1$ \hmpc\ and $r\gtrsim
3$ \hmpc, we find that the PVD decreases with increasing galaxy
luminosity. This result is mostly driven by the fraction of satellite
galaxies $\fsat$, which is well-constrained by the shape and amplitude
of the correlation function. We find $\fsat \sim 25\%$ for galaxies
fainter than $L_\ast$, while for brighter galaxies the satellite
fraction rapidly declines, creating the decrease in the PVD with
luminosity. At $r=1$ \hmpc, the PVD has no dependence on luminosity
because satellite galaxies dominate the statistics for both bright and
faint objects. Recent measurements of the PVD in Fourier space using
the ``dispersion model'' have reported a strong decline in PVD with
increasing luminosity at $k=1$ \mpch. We test this method with our HOD
models, finding that the dispersion model roughly recovers the shape
of PVD with scale, but that there is no consistent comparison between
the PVD at a given $k$-scale and the true dispersion at a given value
of $r$. This results in a luminosity dependence in $k$-space that is
stronger than in configuration space. The luminosity dependence of the
HOD results in Fourier space are consistent with those measured at
$k=1$ \mpch; thus the recent measurements of the PVD are fully
explainable in the context of halo occupation models. The
normalization of the PVD is lower than predicted by our fiducial
model, and reproducing it requires a lower value of $\om$ ($\sim 0.2$
instead of 0.3), a lower value of $\s8$ ($\sim 0.7$ instead of 0.9),
or a significant bias between the velocity dispersion of galaxies and
dark matter in massive halos.

\end{abstract}

\section{Introduction}

Galaxy peculiar velocities, their motions with respect to the smooth
Hubble flow, offer a unique window into both cosmology and galaxy
formation. The galaxy velocity field reflects the matter density of
the universe, $\om$, and the amplitude of matter fluctuations
(\citealt{peebles:76, sargent_turner:77, kaiser:87}). This is true at
large scales, where the infall velocities of galaxies toward overdense
regions are determined by the overall amount of matter and its
distribution at large scales, and it is also true at small scales,
where the random motions of galaxies of galaxies are determined by the
shape and normalization of the mass function of dark matter halos
(\citealt{sheth:96}). But peculiar velocities also provide information
about galaxy formation and evolution; just as galaxies are a biased
tracer of the spatial distribution of dark matter, they can also be
biased with respect to the velocity distribution of dark matter
(\citealt{berlind_etal:03, yoshikawa_etal:03,
  faltenbacher_etal:05}). Pairwise velocity statistics generally
differ even if galaxies have the same peculiar velocities as local
dark matter particles because spatial bias changes the weighting of
those velocities. This is of particular importance at small scales,
reflecting the accretion histories and evolution of galaxies within
groups and clusters. A full model of galaxy formation must reproduce
both the real-space clustering of the observed galaxy distribution and
the distribution of velocities, and the dependence of these
distributions on luminosity and galaxy type.

In this paper we make predictions for the pairwise velocity dispersion
of galaxies (hereafter PVD, or $\sv$) based on the observed real-space
clustering of galaxies in the Two-Degree Field Galaxy Redshift Survey
(2dFGRS; \citealt{colless_etal:01}). We use the Halo Occupation
Distribution (HOD) modeling method, in which the relation between
galaxies and mass is described at the level of virialized dark matter
halos (\citealt{kauffmann_etal:97, jing_etal:98, benson_etal:00, seljak:00,
  peacock_smith:00, ma_fry:00, roman_etal:01,
  berlind_weinberg:02}). The mass function and spatial clustering of
dark matter halos themselves, for given cosmological parameters, are
well-determined from numerous numerical and analytic studies (e.g.,
\citealt{sheth_tormen:99, jenkins_etal:01, sheth_mo_tormen:01,
  seljak_warren:04, tinker_etal:05a, warren_etal:05}). In our
approach, the parameters that determine the relationship between halo
mass and the average number of galaxies within a halo are constrained
by observational measurements of the projected two-point correlation
function \wp\ (see, e.g., \citealt{zheng:04, zehavi_etal:04,
  zehavi_etal:05, tinker_etal:05a}). This technique is similar to the
conditional luminosity function (CLF) analysis of
\cite{yang_etal:03}, but utilizes spatial clustering over a wide
range of scales, linear to non-linear. Once the parameters of the
occupation function have been determined, predictions can be made for
other clustering statistics either analytically or through the
creation of mock galaxy distributions with numerical simulations. In
this paper we obtain the PVD by populating the halos of a
high-resolution collisionless N-body simulation with galaxies such
that both the spatial clustering as a function of luminosity and the
galaxy luminosity function match observations. This approach
complements the analytic model for the PVD of \cite{slosar_etal:05},
and our qualitative conclusions about the PVD are consistent with
their analysis.

Directly measuring peculiar velocities requires requires knowledge of
the true distance to galaxies, thereby removing the Hubble flow from
the measured redshift. This limitation subjects these measurements to
shot noise and sample variance. To circumvent this problem, many
studies have used measurements of the two-point galaxy correlation
function in redshift space $\xirp$, which is distorted along the line of
sight by the peculiar velocities (e.g., \citealt{davis_peebles:83,
  bean_etal:83}). In this method, the galaxy PVD is obtained by
modeling the redshift-space correlation function at a given projected
separation $r_p$, assuming that the PVD is a constant along the line
of sight. This method has been used in more recent measurements of the
galaxy PVD by \cite{jing_etal:98}, \cite{zehavi_etal:02}, and
\cite{hawkins_etal:03}. We will compare our predictions to the 2dFGRS
results of \cite{hawkins_etal:03}, demonstrating qualitative agreement
between the measurements and a model normalized to a matter density of
$\om=0.2$. A quantitative comparison is difficult due to the
flux-limited sample used in \cite{hawkins_etal:03}, and to possible
systematic errors of the model used to infer the PVD from the observed
$\xirp$.

Recently, Jing \& B\"orner (\citeyear{jing_borner:04}, hereafter JB)
and Li et.\ al.\ (\citeyear{li_etal:05}, hereafter L05) have presented
measurements of the PVD obtained by modeling observations of the
redshift-space correlation function for volume-limited samples in the
2dFGRS (JB) and Sloan Digital Sky Survey (L05). One notable result of
these two studies is the luminosity dependence of the PVD at small
scales. Both groups measure a clear trend of decreasing velocity
dispersion with increasing luminosity, until approximately
$M_\star-1$, whereupon the PVD rapidly rises. JB claim that this
luminosity dependence is counterintuitive, and both papers show it is
not reproduced in a halo occupation model.

The technique of JB and L05 is to infer the PVD from redshift-space
clustering in Fourier space using the ``dispersion model'' of the
redshift-space power spectrum (e.g., \citealt{cfw:94, cfw:95,
  peacock_dodds:94, hatton_cole:98, hatton_cole:99}; see
\citealt{roman:04} for a thorough review and critique of this model).
In their implementation of the dispersion model, JB and L05 use the
redshift-space power spectrum at a fixed $k$-scale and fit the data as
a function of angle, as opposed to a constant projected
separation. The meaning of the velocity dispersion obtained in this
case is less clear. JB argue that the qualitative transformation
between the PVD in Fourier space and that in configuration space
should be $k \leftrightarrow 1/r$, an inference motivated by the
transition between one-halo and two-halo clustering, which appears to
occur at 1 \mpch\ in $k$-space and 1 \hmpc\ in configuration
space. The $k \leftrightarrow 1/r$ transformation is also supported by
tests with mock galaxy distributions (\citealt{jing_borner:01b}). JB
and L05 compare their results to an implementation of a CLF fit to the
2dFGRS of \cite{yang_etal:03}. This CLF model predicts a PVD that
increases with luminosity, nearly orthogonal to the observed trend.

In this paper we test the JB implementation of the dispersion model
against our HOD models. We find that the fourier space dispersion
measured by JB on $k=1$ \mpch\ can, in its interpretation as a
configuration space dispersion on $r = 1$ \hmpc, be systematically off
the true configuration spcae value at the $20-40\%$ level, and only
roughly follows the true dependence of the PVD with configuration
space scale. We pay particular attention to the question of what
physical scale the $k=1$ \mpch\ measurements are probing. In
configuration space, our HOD models predict that $\sv$ should be
nearly independent of luminosity at $r=1$ \hmpc. At smaller and larger
scales, $\sv$ has a weak dependence on luminosity, such that brighter
galaxies have lower dispersions. Applying the dispersion model
technique to our mock galaxy distributions, we demonstrate that the
$k=1/r$ comparison is reasonable for only a small set of luminosity
samples, and in general there is no consistent comparison between
dispersion model results and the true PVD at a given scale. Thus in
Fourier space the dependence of the PVD on galaxy luminosity is
steeper than that in configuration space, leading to the strong trend
measured in the JB and L05 results.

\section{Data and Modeling}

\subsection{2dFGRS Data}

We make predictions for the PVD and its dependence on luminosity by
inferring the mean halo occupation function from measurements of the
projected galaxy correlation function \wp\ of the 2dFGRS. These
measurements are similar to those presented in \cite{norberg_etal:01}
and \cite{norberg_etal:02}, who measured \wp\ from a sample of $\sim
160,000$ redshifts, but our data have been updated to include the full
data release of the 2dFGRS, which contains $\sim 221,000$
redshifts. Further details about the survey can be found in
\cite{colless_etal:01, colless_etal:03} and \cite{norberg_lumfunc:02}.
 

\begin{figure}
\centerline{\epsfxsize=3.5truein\epsffile{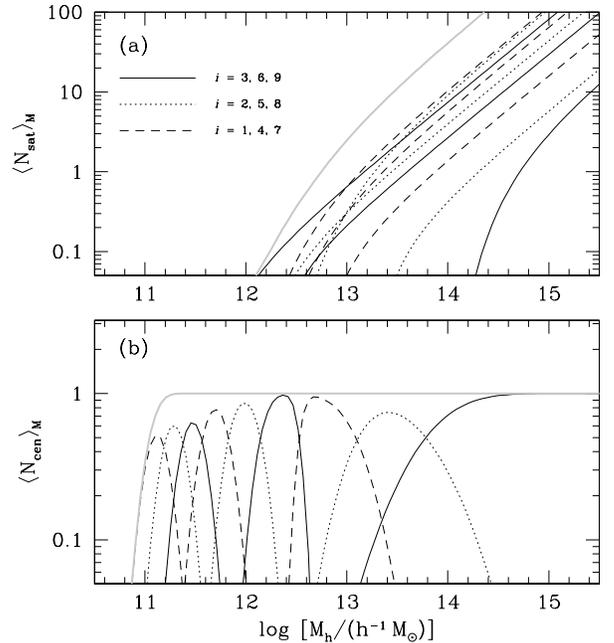}}
\caption{ \label{hod} Panel (a): Satellite occupation
  functions for all nine magnitude bins, as listed in Table 1. The
  curves go in order of faintest to brightest galaxies from left
  to right. The thick gray line shows the total satellite occupation
  function for all galaxies. Panel (b): Central occupation functions
  for the same magnitude bins. The ordering of the lines is the same
  as in panel (a). The thick gray line is the sum of all central
  occupation functions. }
\end{figure}

The details of the clustering measurements can be found in Norberg et
al.~(in preparation). We present here a brief summary of the
calculations. Using the full 2dFGRS survey we create volume limited
samples, with faint limits from $\mbj = -17.0$ to $\mbj = -21.0$, each
sample 0.5 magnitudes\footnote{All magnitudes quoted in this paper
  assume $h=H_0/100=1$. All magnitude bins will be referred to by
  their lower limit or by a reference number $i$, as listed in Table
  1.} wide, for a total of nine samples. All galaxies brighter than
$\mbj = -21$ are grouped into a single sample. As in
\cite{norberg_etal:01, norberg_etal:02}, a careful account of the
selection function is made and the correlation functions are obtained
using the standard \cite{landy_szalay:93} and \cite{hamilton:93}
estimators, with typically 100 times more randoms than galaxies. The
projected correlation function is estimated by integrating
$\xi(r_p,r_\pi)$ out to $r_{\pi,{\rm max}} = 70$~\hmpc, providing a
stable estimate for \wp\ out to at least $r_p = 40$~\hmpc. Due to the
sensitivity of the results on close pair incompleteness, clustering
results from 2dFGRS are only really reliable beyond typically $\sim
150$~\hkpc, which is the limit adopted in this paper too. The
correlation function is measured in thirteen radial bins, spaced
evenly by 0.2 in $\log_{10}r$ beginning at $\log_{10}r = -0.7$.

The errors on the clustering measurements are estimated by a bootstrap
resampling technique on roughly equal sized subregions, of which there
are 16 in total (8 in each 2dFGRS region, covering in each region
approximatively the same survey area). The main reason for choosing
this technique over mock surveys is that it allows error estimates to
be obtained for samples with different clustering properties, whereas
current available mocks need to be tuned for each sample.  Moreover
the existence of large structures in the survey (see, e.g.,
\citealt{baugh_etal:04}) are difficult to account for in the available
mocks, but with bootstrap resampling of large subregions this type of
cosmic variance can, to some extent, be account for. We create 100
bootstrap resamplings, which is sufficient to obtain a stable error
covariance matrix for each subsample (see
\citealt{porciani_norberg:06} for details on stability tests for error
matrices from bootstrap resamplings). We note though that this number
of bootstrap resamplings is not large enough to address the covariance
between the subsamples. For that reason, in this paper, we have to
ignore the existing covariance between `neighboring' volume limited
samples. We use principal component analysis of the covariance matrix
to calculate the $\chi^2$ of a given HOD model (e.g.,
\citealt{porciani_giavalisco:02}), using the first twelve principal
components.

\subsection{HOD Models}

To model \wp, we use the analytic model described in detail in
\cite{tinker_etal:05a} (see also, \citealt{berlind_weinberg:02,
  cooray_sheth:02, zheng:04}). The mean occupation function is divided
into two components; galaxies located at the center of mass of the
host halo, $\ncen$, and satellite galaxies distributed throughout the
host halo with (usually) the same distribution as the dark matter,
$\nsat$. For galaxy samples defined by a luminosity or magnitude
threshold, $\ncen$ can be defined by a single halo mass scale below
which halos are not able to contain galaxies in the sample. For
magnitude bin samples, $\ncen$ must have both a minimum and a maximum
mass scale, the latter being defined by the minimum mass of the next
brightest sample. A simple example would be to use a square window as
the central occupation function, such that $\nceni=1$ for $\mmin^i <
M_h < \mmin^{i+1}$, where $i$ denotes magnitude bin. This
parameterization assumes a unique mapping of central galaxy luminosity
onto halo mass. A more physical model takes into account the scatter
between mass and luminosity, such as that used by \cite{zheng_etal:05}
to model the occupation functions of semi-analytic and hydrodynamic
simulations of galaxy formation. Here we modify their $\ncen$ to take
into account binned magnitude samples, i.e.

\begin{eqnarray}
\label{e.ncen}
\nceni & = & \frac{1}{2}\left[ 1+\mbox{erf}\left(\frac{\log M - \log \mmini}{\sigmaMi} \right)
	\right] - \ncenip, \,\,\, 1\le i\le 8, \nonumber \\ 
\nceni & = & \frac{1}{2}\left[ 1+\mbox{erf}\left(\frac{\log M - \log \mmini}{\sigmaMi} \right)
	\right], \,\,\, i=9,
\end{eqnarray}

\noindent where $\nceni$ represents central galaxies in magnitude bin
$i$, $\mmini$ is the cutoff mass scale for central galaxies,
$\sigmaMi$ controls the width of the cutoff mass range, and ${\tt
  erf}$ is the error function. For luminosity threshold samples,
$\mmin$ is defined as the mass at which $\ncen=0.5$, but in equation
(\ref{e.ncen}) this mass can differ from $\mmini$. The form of
equation (\ref{e.ncen}) guarantees that the sum of $\nceni$ over all
$i$ will never be larger than one, which would be unphysical. The form
we use for the satellite galaxy occupation function is

\begin{equation}
\label{e.nsat}
\nsati = \exp \left(-\frac{\mcuti}{M-\mmini}\right) \left( \frac{M}{M_1^i}\right)^{\asat},
\end{equation}

\noindent where $\mcuti$ is the cutoff mass scale at which $\nsati$
transitions between a linear dependence of $N$ on $M$ and an
exponential cutoff, $\monei$ is the amplitude of the occupation
function (for $M_1^i \gg \mcuti$, $\monei$ is the mass at which halos
host on average one satellite of magnitude $i$), and $\asat$ is the
power-law dependence of $\nsati$ on host mass. We assume the scatter
about the mean follows a Poisson distribution (\citealt{zheng_etal:05,
  kravtsov_etal:04}). Although $\asat$ is a free parameter, we assume
that it is constant for all magnitude bins. Allowing $\mcuti$ to vary
as a function of magnitude allows enough flexibility in the satellite
occupation function to model \wp\ for each bin.

For spatial clustering, differentiating between central and satellite
galaxies is important because pairs of galaxies that involve a central
and a satellite galaxy have a different distribution in the
small-scale (i.e.\ `one-halo') regime than satellite-satellite
pairs. This differentiation is also important when modeling galaxy
velocities. Central galaxies are assumed to be at rest with the center
of mass of the halo, while satellites have random motions roughly
equal to the virial dispersion of the dark matter halo. Thus, pairs of
two satellite galaxies in a common halo have a dispersion $\sqrt{2}$
times larger than pairs of central and satellite galaxies. For
two-halo pairs, in which each galaxy comes from a separate halo, the
pairwise velocity of the halos contributes to all pairs, but the
internal halo dispersions make an additional contribution to pars
involving satellite galaxies.  The ratio of satellites to total
galaxies is therefore important in determining the behavior of the PVD
as a function of luminosity. Although the assumption that central
galaxies have no random motions is a working approximation, in tests
we find that the results of this assumption are nearly identical to
models in which central galaxies have velocities in agreement with
recent observational results of \cite{vdb_etal:05}, who infer
$\sigma_{\rm cen} \approx 0.25 \sigma_{\rm sat}$. For pairwise
statistics these dispersions add in quadrature, thus central motions
of this magnitude only increase the central-satellite dispersion by
3\%.

To calculate \wp\ for a given set of HOD parameters, we must first
assume a cosmology and a form of the linear matter power
spectrum. Because we will use the results of our fitting procedure to
populate the halos identified in a collisionless N-body simulation, we
take the same values as those used in the simulation. We assume a flat
\lcdm\ model with $\om=0.3$ and a linear amplitude of fluctuations of
$\s8=0.9$, where $\s8$ is the variance of matter fluctuations on an 8
\hmpc\ scale. The shape of the linear matter power spectrum was
computed using \cmbfast\ (\citealt{cmbfast}), with $h=0.7$ and
$\omb=0.04$. The simulation was performed using the Hashed Oct-Tree
code of \cite{warren_salmon:93}. The simulation box is 400 \hmpc\ per side,
with a total of $1280^3$ particles, giving a mass resolution of
$2.5\times 10^9$ \hmsol.  Halos are identified in the simulation by
the friends-of-friends technique (\citealt{davis_etal:85}) with a linking
length 0.2 times the mean interparticle separation. For the
lowest luminosity, $\mbj\sim -17$ galaxies, the minimum mass roughly
corresponds to a 50-particle halo.

To calculate the model \wp\ for each bin, the HOD parameters are set
to match the abundance of galaxies within each magnitude bin $\ngavgi$
as determined by the $b_J$ luminosity function of
\cite{norberg_lumfunc:02}, with Schechter function parameters
$\Phi_\ast = 0.0161$, $\mstar = -19.7$ and $\alpha = -1.21$. This is
done by calculating the value of $\mmini$ required to match $\ngavgi$
once the other parameters for bin $i$ have been chosen. For
high-luminosity bins ($i\ge 7$, or $\mbj\le -20$, which we will refer
to as the `bright' samples, while referring to $i<7$ as the `faint'
samples), we allow $\slogmi$ to be a free parameter. Recall that the
shape of the cutoff in $\nceni$ physically represents the scatter
between halo mass and central galaxy luminosity; at bright magnitudes
this scatter is expected to be larger due to the steepness of the halo
mass function in this regime and the errors in 2dFGRS galaxy
magnitudes. For faint samples the scatter is expected to be smaller
and magnitude errors do not increase it. We therefore fix $\slogmi$ to
be 0.15 for these samples. Leaving $\slogmi$ as a free parameter for
these samples does not result in a statistically better fit to \wp,
and in tests we find that the results for the PVD are not sensitive to
the value chosen.


\begin{figure*}
\centerline{\epsfxsize=5.5truein\epsffile{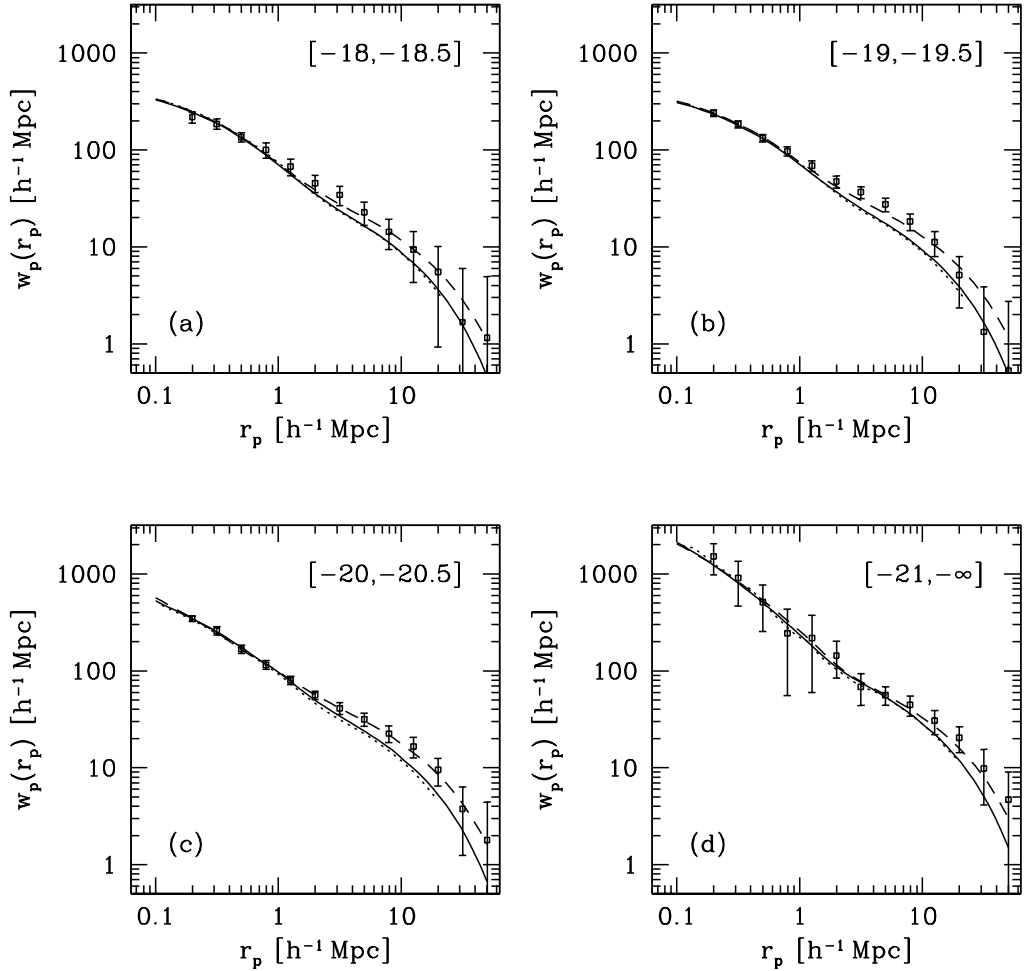}}
\caption{ \label{xi} Panel (a): Model and observed correlation
  functions for the galaxies with $-18 \ge \mbj \ge -18.5$ ($i=3$ in
  Table 1). Points with error bars show the measured \wp\ from the
  full 2dFGRS data release. The solid line plots the best-fit analytic
  model, as listed in Table 1. The dotted line shows the \wp\ obtained
  from the numerical simulation, once the halos have been populated
  with the same HOD parameters. The dashed line shows the best-fit HOD when
  the parameters of \cite{sanchez_etal:05} are used to calculate the
  linear matter power spectrum. Panels (b)---(c) are the same as (a),
  but for magnitude bins $i=5$, 7, and 9. }
\end{figure*}

We find the best-fitting HOD models using the Markov Chain Monte Carlo
(MCMC) technique (see, e.g., \citealt{doran_muller:04}). In MCMC, an
HOD parameter set is randomly chosen from a Gaussian distribution
centered on the last accepted parameter set in the MCMC chain. Each
element in the chain contains the full set of HOD parameters for all
nine magnitude bins. A model is automatically accepted into the chain
if it results in a lower $\chi^2$ value than the previously accepted
model, and accepted with the probability $p = \exp(-\Delta \chi^2/2)$
if the $\chi^2$ is higher. The number of parameters in our model is
22, consisting of two parameters ($\monei$ and $\mcuti$) for each
$i=1-6$, three parameters for each $i=7-9$ ($\monei$, $\mcuti$, and
$\slogmi$), and $\asat$, which is the same for all $i$. Because the
values of $\mmini$ are set by the galaxy number densities, they are
not free parameters
\footnote{We neglect the errors on the galaxy number densities, which
  are of order a few percent. Due to the steepness of the halo mass
  function, these errors have little effect on $\mmini$. The dominant
  source of uncertainty are the \wp\ measurements.}.  The total
$\chi^2$ for each model is the sum of the $\chi^2$ values obtained for
the \wp\ in each magnitude bin. This large number of parameters makes
MCMC an ideal technique for exploring the likelihood of possible
solutions and of predictions these solutions make for the
luminosity-dependence of the PVD. To check the chain for convergence
we use the power spectrum method of \cite{dunkley_etal:05}. The
element with the minimum $\chi^2$ is taken to be our fiducial model,
which is listed in Table 1. Although we list the $\chi^2$ value for
each \wp\ in Table 1, it should be noted that this model was chosen by
its total $\chi^2$. Therefore, for each individual bin $i$ there will
be models for which $\chi^2_i$ is lower than that listed in Table
1. We select a subset of 1,000 models from the chain with which to
test the constraints on HOD parameters and the PVD inferred from
them. Table 1 also lists the galaxy bias factors $b_g^i$ of the
fiducial simulation, as well as the dispersion in $b_g^i$ from the
1,000 MCMC models. The galaxy bias is straightforward to calculate
analytically once the HOD parameters are known, and is determined by

\begin{equation}
\label{e.bias}
b_g^i = \frac{1}{\ngavgi}\,\int_0^\infty dM \, \frac{dn}{dM} \, \navgi \, b(M), 
\end{equation}

\noindent where $dn/dM$ is the halo mass function
(\citealt{jenkins_etal:01}) and $b(M)$ is the halo bias function
(\citealt{tinker_etal:05a}). The variance of $b_g^i$ is $\sim 1-2\%$
for most bins, with the tightest constraints for bins $i=4-7$ ($-18.5
\ge \mbj \ge -20.0$). For lower luminosity bins the limited volume of
the samples increases the errors on \wp, while for higher luminosity
bins the low number density of galaxies creates significant shot noise
in \wp. It should be stressed that the errors in $b_g^i$ are for a
given cosmological model. The galaxy bias obtained from \wp\ is
degenerate with the assumed linear matter power spectrum $\plin$.

Figure \ref{hod} shows the occupation functions from the fiducial
model for all nine magnitude bins. Panel (a) plots, from left to
right, $\nsati$ as a function of $\log M_h$ for $i=1-9$. The thick
grey line is the sum of all $\nsati$, which effectively represents the
satellite occupation function for a magnitude threshold sample for
galaxies brighter than $\mbj=-17$. As expected from both the shape of
the galaxy luminosity function and from the increased relative bias as
a function of luminosity (see, e.g., \citealt{zehavi_etal:05,
  norberg_etal:02}), the amplitude of the satellite occupation
function $\monei$ monotonically increases with the magnitude of the
bin. The best-fit values of $\mcuti$ are not monotonic with
luminosity, and for $i=1$ and 2 the satellite cutoff occurs at higher
mass than $i=3$. Due to the larger errors on \wp\ for lower luminosity
bins, and the fact that $\nsati$ is not well-constrained for
occupation numbers significantly less than unity, there is a large
range of $\mcuti$ allowed by the data for these magnitude bins. We
will discuss this further in the following section.

Figure \ref{hod}b plots $\nceni$ as a function of $\log M_h$ for the
same magnitude bins. The thick grey line, once again showing the sum
of all nine $\nceni$ functions, resembles the central occupation
function of a magnitude threshold galaxy sample. For the brighter
samples, where $\slogmi$ is allowed to vary, the shape of the cutoff
becomes progressively softer with magnitude. This is expected for two
reasons. First, the halo mass function contains an exponential cutoff
above $M_h \sim 10^{13}$ \hmsol, and this steepness is expected to
increase the physical scatter between halo mass and central galaxy
luminosity. HOD models of the brightest SDSS galaxies by
\cite{zehavi_etal:05} that utilized an exponential cutoff were
statistically preferred to models with a step-function form of
$\ncen$. Second, the magnitude errors in the 2dFGRS are significantly
larger than those of SDSS measurements. The exponential cutoff in the
galaxy luminosity function creates asymmetric scatter, as galaxies of
lower luminosity are scattered into a brighter bin more often than
higher luminosity galaxies are scattered down.


\begin{deluxetable*}{rrrccccccc}
\tablecolumns{10} 
\tablewidth{37pc} 
\tablecaption{HOD Parameters for Models in Figure \ref{hod}}
\tablehead{\colhead{$i$} & \colhead{$\mbj$} & \colhead{$\chi^2$} & 
\colhead{$\mmin$} & \colhead{$M_{\rm cen}^{\rm max}$} & \colhead{$M_1$} & \colhead{$\mcut$} & \colhead{$\slogm$} &
\colhead{$b_g$} & \colhead{$\delta_{b_g}$} }
\startdata

1 & -17.0 & 15.6 & 1.11$\times 10^{11}$ & $1.27\times 10^{11}$ &  1.01$\times 10^{13}$ & 4.17$\times 10^{12}$ &  0.15 & 0.90 & 0.02\\  
2 & -17.5 & 6.5 & 1.57$\times 10^{11}$ & $2.01\times 10^{11}$ & 1.08$\times 10^{13}$ & 1.07$\times 10^{13}$ &  0.15 & 0.92 & 0.03\\   
3 & -18.0 & 6.3 & 2.38$\times 10^{11}$ & $2.86\times 10^{11}$ & 1.44$\times 10^{13}$ & 5.71$\times 10^{11}$ &  0.15 & 0.94 & 0.015\\   
4 & -18.5 & 10.9 & 3.71$\times 10^{11}$ & $5.12\times 10^{11}$ & 1.80$\times 10^{13}$ & 5.64$\times 10^{12}$ &  0.15 & 0.93 & 0.011\\  
5 & -19.0 & 13.1 & 6.73$\times 10^{11}$ & $9.15\times 10^{11}$ & 2.61$\times 10^{13}$ & 1.84$\times 10^{12}$ &  0.15 & 0.96 & 0.011\\  
6 & -19.5 & 12.5 & 1.40$\times 10^{12}$ & $2.32\times 10^{12}$ & 3.83$\times 10^{13}$ & 1.45$\times 10^{12}$ &  0.15 & 1.03 & 0.007\\  
7 & -20.0 & 9.7 & 3.37$\times 10^{12}$ & $4.66\times 10^{12}$ & 6.93$\times 10^{13}$ & 6.49$\times 10^{12}$ &  0.09 & 1.13 & 0.009\\   
8 & -20.5 & 25.5 & 1.22$\times 10^{13}$ & $2.67\times 10^{13}$ & 1.82$\times 10^{14}$ & 2.19$\times 10^{13}$ &  0.33 & 1.29 & 0.03\\  
9 & -21.0 & 5.2 & 6.05$\times 10^{13}$ & --- & 2.49$\times 10^{14}$ & 3.48$\times 10^{14}$ &  0.56 & 1.69 & 0.08\\   

\enddata \tablecomments{All masses are in units \hmsol. The best-fit
  value of $\asat$ from equation (\ref{e.nsat}) is 1.03. $M_{\rm
    cen}^{\rm max}$ is the mass at which $\nceni$ is maximum. $b_g$ is
  the galaxy bias factor, and $\delta_{b_g}$ is the dispersion of bias
  values from the MCMC chain.}
\end{deluxetable*}

\subsection{Mock Galaxy Distributions}

To create mock galaxy distributions, we populate the identified halos
in the N-body simulation. The number of satellite galaxies for each
halo in each magnitude bin is chosen from a Poisson distribution with
mean $\nsati$. The presence of a central galaxy is determined from a
nearest integer distribution from the sum of all $\nceni$, and the
luminosity of the central galaxy is chosen from the relative
distribution of $\nceni$.  The central galaxies are placed at the
center of mass of the halo, while the satellite galaxies are placed
randomly throughout the halo according to the universal density
profile of \cite{nfw:97}.  For each halo mass, the appropriate
parameters of the \cite{nfw:97} profile are calculated for our assumed
cosmology from the method of \cite{bullock_etal:01}. The number
density profiles of the satellite galaxies are assumed to follow the
density profile of the dark matter. As stated above, the central
galaxy is given the velocity of the center of mass of the halo, while
the satellite galaxies are given an additional random velocity in each
direction drawn from a Gaussian distribution of width

\begin{equation}
\label{e.sigmah}
\sigma_h^2 = \frac{GM_h}{2R_{200}},
\end{equation}

\noindent where $R_{200}$ is the radius at which the mean interior
density of the halo is 200 times the universal mean, a value that
correlates well on average with the mean densities of
friends-of-friends halos with a 0.2 linking length. The relation
between $M_h$ and $R_{200}$, $R_{200} = (3M_h/4\pi 200\rho_m)^{1/3}$,
makes $\sigma_h \propto M_h^{1/3}$. Equation (\ref{e.sigmah}) assumes
no velocity bias for satellite galaxies. A more general form of equation
(\ref{e.sigmah}) would have a constant of proportionality $\alpha_v^2$
relating satellite motions to dark matter
(\citealt{berlind_weinberg:02, tinker_etal:05b}).

Figure \ref{xi} shows the \wp\ data and the HOD models for four of the
magnitude bins, as labeled in each panel. The points with error bars
are the 2dFGRS measurements, the solid lines are the fiducial analytic
model, and the dotted lines show the measured correlation functions
from the populated N-body simulation. The total $\chi^2$ of this model
is 105.3 for $9\times 12 - 22 = 86$ degrees of freedom, yielding a
$\chi^2$ per degree of freedom of 1.22. But from inspection of Figure
\ref{xi} it is apparent that the cosmology chosen lacks large-scale
power relative to the data. If these data are fit using a $\plin$
consistent with the recent analysis of the 2dFGRS large-scale galaxy
power spectrum and CMB results (\citealt{sanchez_etal:05}), the
difference at large scales is ameliorated, and the total $\chi^2$ is
reduced. These fits are shown with the dashed lines in Figure
\ref{xi}. The fits at small scales are unchanged, but at large scales
the models match the amplitude of the observations. The $\chi^2$ per
degree of freedom for these fits is 0.61, a value that suggests the
possibility that the errors on the data have been overestimated,
perhaps by neglecting the covariance in \wp\ between luminosity bins,
or by using too many principal components in the error analysis
(\citealt{porciani_norberg:06}).  We use the standard
\lcdm\ fits in order to populate the simulation. Using a simulation
with a \cite{sanchez_etal:05} power spectrum would likely produce
small differences in the predicted PVD, but it would not radically
change our predictions because the best-fit halo occupation functions
are not very sensitive to the assumed power spectrum shape.

\subsection{Satellite Fraction of Galaxies}

A key quantity for interpreting the results of the HOD analysis is the
satellite fraction $\fsat$, determined for each bin from the HOD
parameters by

\begin{equation}
\label{e.fsat}
\fsati = \frac{1}{\ngavgi}\int_0^\infty\, dM\, \frac{dn}{dM}\, \nsati .
\end{equation}

\begin{figure}
\centerline{\epsfxsize=3.5truein\epsffile{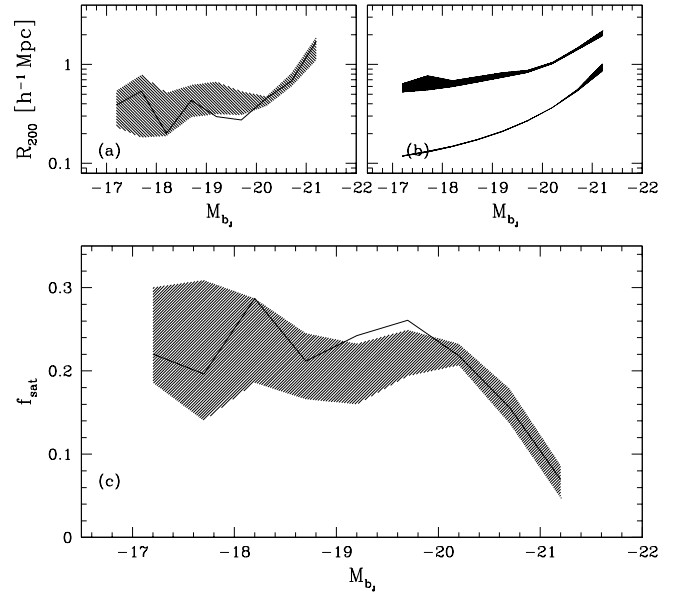}}
\caption{ \label{fsat} Panel (a): Radii of $\mcuti$ halos, below which
  the satellite occupation function is exponentially suppressed, as a
  function of magnitude. The thick solid line is the relation from the
  fiducial model, while the shaded region represents the inner 68\% of
  the distribution of $\mcuti$ values about the median, taken from the
  MCMC chain. Panel (b): Radii of $\mmini$ halos (bottom shaded
  region), the cutoff scale for central galaxies, and $\mONEi$ halos
  (upper shaded region), which on average contain one satellite. These
  results are taken from the MCMC analysis only. Panel (c): Fraction
  of galaxies that are satellites as a function of magnitude. Line and
  shaded region are as in panel (a). }
\end{figure}

\noindent We plot this quantity as well as the different mass scales
from the occupation functions in Figure \ref{fsat}. Panel (a) plots
the radii of $\mcuti$ halos as a function of magnitude for both the
fiducial model and the MCMC chain. We present this quantity as a
radius rather than a mass because the physical sizes of halos
influence the scales at which one-halo and two-halo pairs can
contribute. The solid black line represents $R(\mcuti)$ for the
fiducial model. The shaded region represents the inner 68\% of the
distribution of $R(\mcuti)$ about the median value from the MCMC
chain. The distribution of $R(\mcuti)$ is calculated for each bin
individually, so the upper and lower bounds on the shaded region do
not represent a single element in the chain, but the shaded region can
be taken as an estimate of the error in $\mcuti$. The bin-to-bin
variation seen in the fiducial model is common for individual
models from the chain. The midpoint of the shaded area is not defined
by a single model, thus the best-fit model need not follow it. The
constraints on $\mcuti$ are weakest at lowest luminosities, where the
observational errors are largest. The shaded region narrows for
$\mbj\le -19.5$ due to the smaller errors on \wp.


Figure \ref{fsat}b shows the results for $\mmini$ and $\mONEi$ for the
MCMC models only, where $\mONEi$ is the mass at which $\nsati = 1$,
which can be larger than $\monei$ due to the exponential cutoff in
equation (\ref{e.nsat}). The shaded regions show the same distribution
about the median as in panel (a), with the lower shaded region
representing the results for $\mmini$ and the upper region
representing $\mONEi$. The constraints on both these parameters are
significantly tighter than those on $\mcuti$. The majority of one-halo
pairs come from halos with $\sim 1$ satellite galaxy, so \wp\ provides
the best constraints on the occupation function around $\mONEi$. The
value of $\mcuti$ influences the shape of $\nsati$ where satellite
occupation is typically less than $\sim 0.5$ and it is therefore
less well constrained by \wp. The value of $\mmini$ is tightly constrained
by both \wp\ and the galaxy number density. Not plotted is the
distribution of $\asat$, which is nearly Gaussian, with a mean of
$\langle \asat \rangle = 1.03$ and a dispersion of 0.03. It should be
noted that this value of $\asat$ is degenerate with the assumed value
of $\s8$; a lower value of $\s8$ would increase the inferred value of
$\asat$ in order to match the observed clustering. 

Figure \ref{fsat}c plots $\fsati$ as a function of magnitude for both
the fiducial model (solid line) and the MCMC chain (shaded region). As
with $\mcuti$, the constraints on $\fsati$ are weaker for fainter
samples, where it is constrained to $\sim 25\%$ about a median value
of approximately 0.24. The median value of $\fsati$ drops gradually to
$\sim 0.20$ for $\mbj \sim \mstar$ galaxies, then falls rapidly for
samples brighter than $-20$, reaching $\fsati = 0.06 \pm 0.02$ for
$i=9$. These trends of the satellite fraction are in good agreement
with the results of \cite{mandelbaum_etal:05}, who derive satellite
fractions from measurements of galaxy-galaxy
lensing. \cite{mandelbaum_etal:05} report $\fsat \approx 30\%$ for
$L<L_\ast$ galaxies (in Sloan $r$-band) and $\fsat \approx 0.13$ for
galaxies significantly brighter than $L_\ast$.

This estimate of $\fsat$ depends only weakly on the
value of $\s8$ assumed. To match the observed \wp\ for a lower value
of $\s8$, $\asat$ increases to compensate for the lack of high-mass
halos. This increases the mean bias of galaxies, but for $\fsat$ the
effects of higher $\asat$ and fewer massive halos nearly cancel. In
tests where we assume $\s8=0.7$, the median $\fsati$ is lower by $\sim
10\%$ than the results in Figure \ref{fsat}, but the 68\% confidence
regions overlap significantly at all magnitudes. The shape of
$\fsati$, particularly the rapid turnover at $\mstar$, is also
preserved.


\begin{figure}
\centerline{\epsfxsize=3.5truein\epsffile{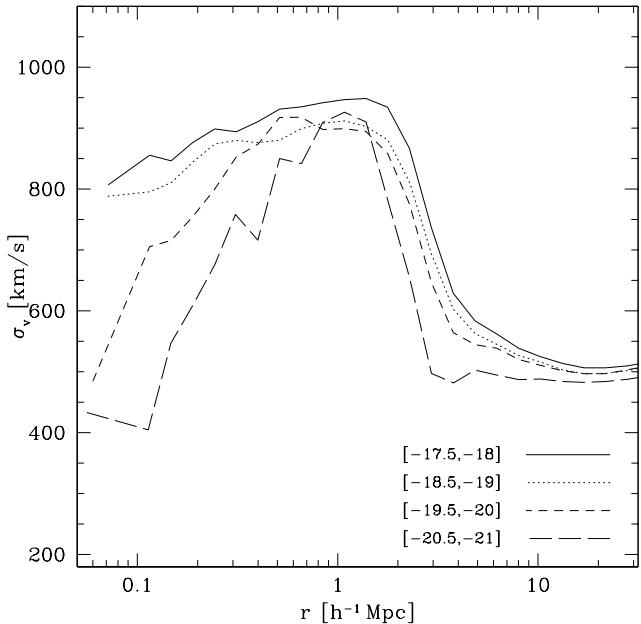}}
\caption{ \label{sigr} Pairwise velocity dispersion as a function of
  radial separation for four magnitude bins, measured from the
  populated simulation. }
\end{figure}

\section{Luminosity and Scale Dependence of the Pairwise Velocity Dispersion}

\subsection{Theoretical Predictions}

We calculate the PVD from our populated simulation by binning pairs in
real space by their radial separation and calculating the
line-of-sight pairwise dispersion as a function of $r$. We focus on
$\sv$ as a function of three-space separation $r$ rather than
projected separation $r_p$ for two reasons: presenting $\sv(r)$ offers
a cleaner picture of the PVD at small scales, and the measurements of
JB and L05 are intended to be of the PVD at fixed scale and not as
function of projected separation. Figure \ref{sigr} shows the radial
dependence of $\sv$ for four bins of increasing magnitude (samples
$i=2,4,6,8$ in Table 1). The general shape of the curves is similar to
those found for dark matter (e.g., \citealt{efstathiou_etal:88}); at
large scales, $\sv$ has little scale dependence, but at $r\sim 3$
\hmpc, $\sv$ rapidly increases towards smaller $r$, peaking at $r\sim
1$ \hmpc. This feature represents the transition between two-halo
pairs, which are dominated by pairs of central galaxies, and one-halo
pairs dominated by pairs of satellite galaxies. At $r<1$ \hmpc, the
PVD turns over and monotonically falls with smaller separation. This
feature reflects two aspects of the one-halo term. At smaller
separation, the relative contribution of central-satellite pairs to
the overall pair count increases. Furthermore, as $r$ decreases,
lower-mass halos begin to contribute one-halo pairs, which further
reduces the PVD. The former effect is more pronounced for brighter
samples because of the smaller fraction of satellite galaxies for
these samples.

At a given separation, the trend of $\sv$ with luminosity can be
inferred from Figure \ref{sigr}. For $r\ge 1$ \hmpc, the dependence of
the PVD on magnitude varies from roughly flat to a moderate decrease
with increasing luminosity. At smaller separations, the dependence
with luminosity becomes stronger. The location of the transition
between between one-halo and two-halo pairs also depends on
luminosity, with lower-luminosity galaxies transitioning to one-halo
pairs at larger separation ($\sim 3$ \hmpc\ for $i=2$) as compared to
bright galaxies ($\sim 1.5$ \hmpc\ for $i=8$). These results are in
good agreement with the analytic HOD calculation of the PVD in
\cite{slosar_etal:05}; both models predict the same characteristic
shape to $\sv(r)$, as well as the same behavior with luminosity at
$r\lesssim 0.5$ \hmpc. There are minor differences between the
predictions, such as the trend of the PVD with luminosity at $r\gtrsim
1$ \hmpc, which in their model increases with increasing luminosity,
which is the opposite of the trend seen in Figure \ref{sigr}. This
discrepancy results from \cite{slosar_etal:05} fixing the satellite
fraction to be a constant value of 0.2 with luminosity. Here we obtain
$\fsat$ directly from the clustering data. \cite{slosar_etal:05}
conclude that the amplitude of the PVD is highly sensitive to the
value of $\fsat$, and we concur.

\begin{figure*}
\centerline{\epsfxsize=5.5truein\epsffile{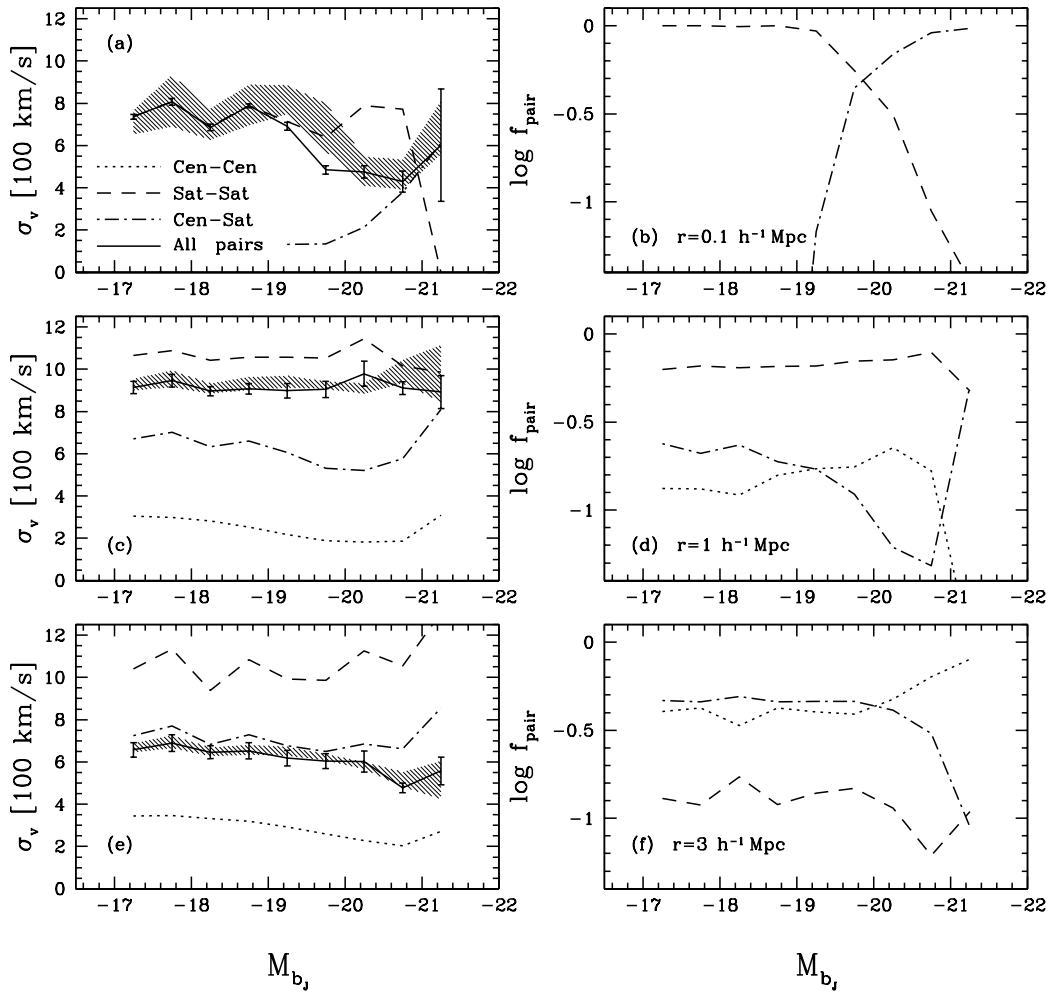}}
\caption{ \label{pair_frac} Panel (a): PVD at $r=0.1$ \hmpc, broken
  down by pair type, as a function of luminosity. The dashed line is
  the PVD for satellite-satellite pairs only, the dash-dotted line
  represents central-satellite pairs, and the dotted line (not seen in
  this panel) is for central-central pairs. The solid line is the PVD
  for all pairs. All curves are obtained from the Fiducial HOD model
  in Table 1. The shaded region is the inner 68\% of the distribution
  of $\sv$ values about the median in each magnitude bin, obtained
  from the MCMC chain. Panel (b): The fraction of the total pair count
  contributed by each type of pair. Line types are the same as in
  panel (a). Panels (c) and (d): PVD and pair fraction for each pair
  type at $r=1$ \hmpc. Line types are as above, with the dotted line
  representing the central-central contribution. Panel (e) and (f):
  Same as above, but for $r=3$ \hmpc. }
\end{figure*}

We use the results of our fiducial model to interpret the luminosity
dependences shown in the previous figure. Figure \ref{pair_frac}
breaks down the relative contributions of the different types of pairs
as a function of magnitude within the radial bins at $0.1$, $1$, and
$3$ \hmpc.  Within isolated halos, galaxy pairs can be either
central-satellite and satellite-satellite. These types of pairs can
also exist between two distinct halos, in addition to the
central-central contribution. For brevity, we refer to these pair
types as cen-sat, sat-sat, and cen-cen in the Figure legend and in the
discussion below. Panel (a) plots $\sv$ for each type of pair at
$r=0.1$ \hmpc, as well as the overall PVD with the solid line. At this
separation, all galaxy pairs are one-halo. The error bars on the solid
line are obtained from jackknife sampling of the simulation volume
into eight octants. Each octant is roughly equal to the volume of the
$\mstar$ sample used here. The shaded region once again shows the
inner 68\% about the median value of $\sv$ at each luminosity from the
MCMC chain. For faint galaxies, $\sv$ is entirely accounted for by the
sat-sat contribution because $\nceni$ and $\nsati$ for these samples
do not overlap in mass; halos massive enough to host satellites are
too massive to have a low luminosity central galaxy. At magnitudes
brighter than $\mbj=-19$, $\sv$ decreases significantly even though
the sat-sat PVD remains constant. For the brightest sample the overall
PVD is equal to the sat-cen dispersion. Panel (b), which plots the
relative contribution of each pair type to the overall pair count as a
function of magnitude, shows why this occurs. Below $\mbj=-19$, sat-sat
pairs account for all pairs. At $\sim \mstar$, the contribution of
cen-sat pairs rapidly increases due both to the presence of central
galaxies and to the lower fraction of satellites. For $\mbj\le -21$,
all pairs involve a central galaxy. For this sample, $\nsati=1$ at
$4\times 10^{14}$ \hmsol. Although the abundance of halos more massive
than this value is low, they do exist in the simulation volume, making
it possible to have sat-sat pairs at this magnitude. But because the
volume of these halos is relatively large, the probability of having
two very bright satellite galaxies within 0.1 \hmpc\ is low.


Figure \ref{pair_frac}c plots $\sv$ for each type of pair at $r=1$
\hmpc. At this separation, which combines both one-halo and two-halo
pairs, there are contributions to $\sv$ from all three pair
types. However, the overall PVD lies between the sat-sat and cen-sat
dispersions, indicating that one-halo statistics still dominate. This
is confirmed in panel (d), which plots the relative fraction of the
pair counts. Sat-sat pairs dominate at all magnitudes until the
brightest bin. For $\mbj\le -21$ galaxies, the contribution of cen-sat
pairs is equal to sat-sat pairs. Figure \ref{fsat}a showed that the
radius of $\mcuti$ halos for this sample is $\gtrsim 1$ \hmpc, roughly
equal to the radial separation shown here. For faint galaxies, cen-sat
pairs are more common than pairs between central galaxies, but the
abundance of cen-sat pairs gradually decreases with magnitude and
cen-cen pairs become more common at $\mstar$, a result of the lower
$\fsat$ values. Panel (e) plots $\sv$ for each pair type at $r=3$
\hmpc. At this separation two-halo pairs dominate the statistics, and
the overall PVD lies between the cen-sat and cen-cen values. In panel
(f), cen-sat and cen-cen pairs are nearly equal for faint galaxies,
but there is once again a transition at $\mstar$ where the
contribution of cen-cen pairs increases and the other pair types
become less common due to the lower satellite fraction.

At small separation, well within the one-halo term, the dependence of
the PVD on galaxy luminosity is strong, with brighter galaxies having
lower dispersions (with the exception of the brightest galaxies, which
live in the highest-mass halos). At intermediate separation, where the
transition between one-halo and two-halo pairs occurs, the dependence
on luminosity is washed out due to the dominance of satellite pairs at
all magnitudes. Outside the one-halo term, but still in the non-linear
regime, there is a weak dependence of $\sv$ on luminosity, falling
from $\sim 700$ \kms\ for $\mbj=-17$ to $\sim 500$ \kms\ for
$\mbj=-20.5$. (From Figure \ref{sigr} it can be seen that this slope
becomes shallower at even larger scales.) At all three scales, the
results of the MCMC chain demonstrate that these predictions are
robust given the errors in the \wp\ measurements. For the latter two
separations, the width of the shaded region is comparable to the
jackknife errors from the simulation itself. At face value, these
results are at odds with the observational results of L05 and JB,
which show a strong dependence of $\sv$ on luminosity at $k=1$ \mpch,
which they argue is equivalent to $r=1$ \hmpc. Our HOD results are
also inconsistent with the CLF prediction shown in both papers, which
shows a strong trend of increasing PVD with luminosity. In section \S
4 we will use the JB technique of obtaining the PVD from the
redshift-space power spectrum to make a proper comparison of our HOD
predictions with these results.


\begin{figure*}
\vspace{-2cm}
\centerline{\epsfxsize=5.0truein\epsffile{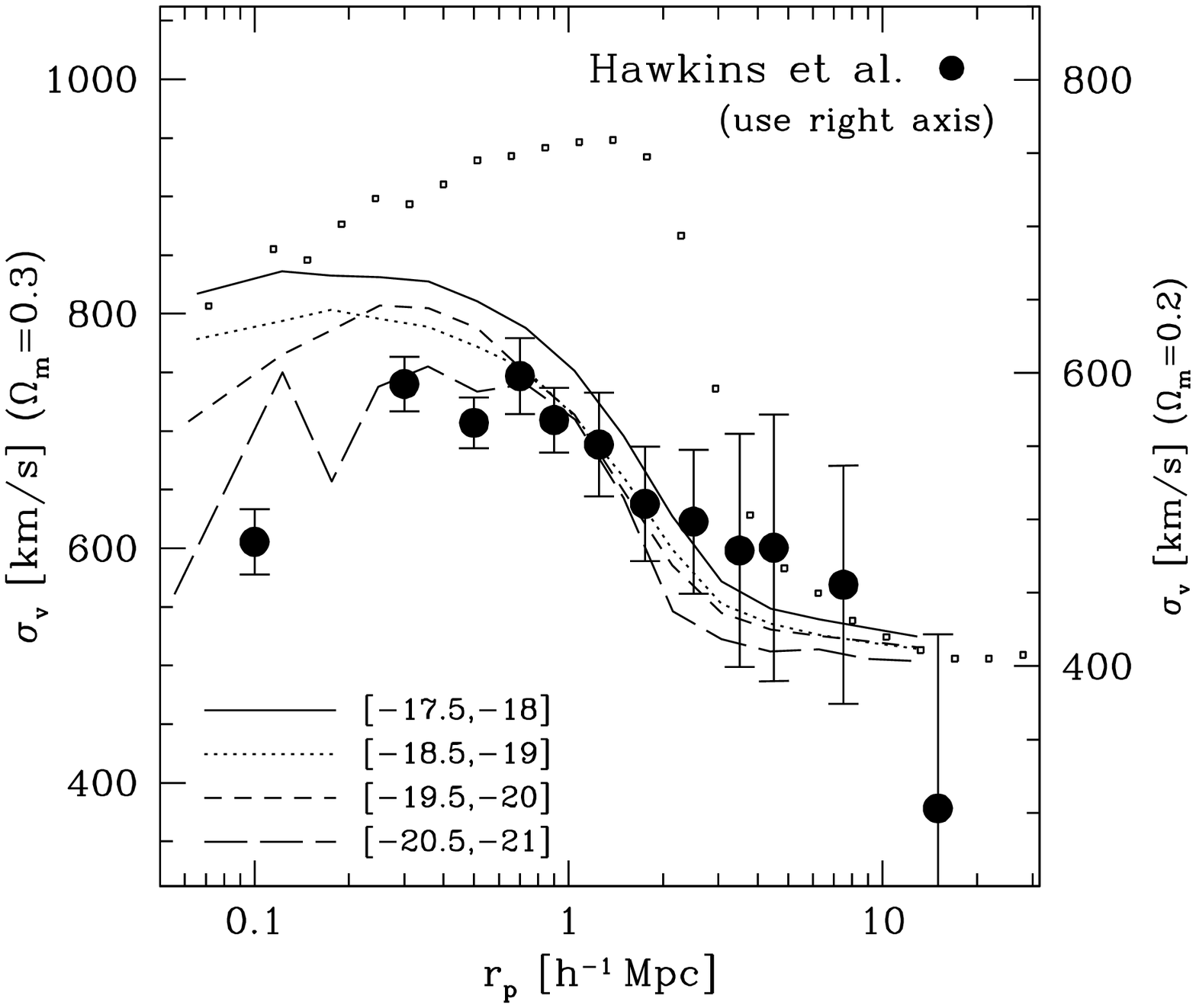}}
\caption{\label{sigrp} Pairwise velocity dispersion as a function of
  projected separation for the same four magnitude bins shown in the
  Figure \ref{sigr} (as a function of 3-d separation). Only pairs with
  a line-of-sight separation less than 40 \hmpc\ are counted. The open
  squares are the results from Figure \ref{sigr} for the faintest bin
  shown ($i=2$). The left-hand $y$-axis is the scale of the PVD as
  measured from the N-body simulation ($\om=0.3$). The right-hand
  $y$-axis is the scale of the PVD for a universe with $\om=0.2$ (see
  text for details). The filled circles show the 2dFGRS measurements
  for a flux-limited sample from \cite{hawkins_etal:03}, which should
  be viewed with respect to the right-hand axis.}
\end{figure*}

\subsection{Comparison with 2dFGRS Measurements}

Estimates of the PVD from the redshift-space correlation function do
not show the shard transition at $r\sim 2$ \hmpc\ seen in Figure
\ref{sigr}. Results from \cite{jing_etal:98}, \cite{zehavi_etal:02},
and \cite{hawkins_etal:03}, from the Las Campanas Redshift Survey, the
SDSS, and the 2dFGRS, respectively, show a smooth increase in PVD from
large to small scales. However, as discussed in \S 1, the PVD is not
measured directly from the data but inferred by model fitting, and
the quantity that is fitted can be interpreted as an ``effective''
velocity dispersion along the line of sight, calculated as a function
of {\it projected} separation $r_p$. In Figure \ref{sigrp}, we plot
the mean line of sight $\sv$ as a function of $r_p$ for pairs with a
real-space line of sight separation of less than 40 \hmpc. The four
curves are for the same magnitude bins as shown in Figure \ref{sigr}. For
comparison, the data from Figure \ref{sigr} for $i=2$ are shown with
the open squares. The sharp transition from one-halo to two-halo pairs
is smoothed out when plotting $\sv(r_p)$ because at a given $r_p$
pairs with larger separations (and lower dispersion) contribute to the
average.

Also shown in Figure \ref{sigrp} are the PVD measurements from
\cite{hawkins_etal:03}. These data follow the axis on the right-hand
side of the plot, the values of which are lower by a factor of
0.8. Thus the amplitude of our HOD prediction is higher than the
data. These calculations assume $\om=0.3$ and no velocity bias. Both
of these assumptions could be wrong, and there exists growing
observational evidence indicating a lower value of $\om$
(\citealt{bahcall_etal:03, vdb_etal:03, cole_etal:05, sanchez_etal:05,
  mohayaee_tully:05, tinker_etal:05a}) and some theoretical evidence
for moderate negative velocity bias (\citealt{klypin_etal:99,
  berlind_etal:03, yoshikawa_etal:03, slosar_etal:05}). Therefore we
consider the overall normalization of our prediction to be somewhat
arbitrary. Read against the right-hand axis, the curves represent
models scaled\footnote{Halo pairwise velocities scale as $\om^{0.6}$
  even in the non-linear regime (\citealt{zheng_etal:02}), while
  virial velocities scale as $\om^{0.5}$, thus to a good approximation
  the pairwise velocities scale as $\om^{0.55}$ at all scales.} by a
factor of $(0.2/0.3)^{0.55} = 0.8$ to depict a universe with
$\om=0.2$. For fixed power spectrum shape, changing $\om$ does not
alter the shape of the halo mass function
(\citealt{zheng_etal:02}). Thus the occupation functions in Figure
\ref{hod}, once scaled to the new value of $\om$, produce identical
fits to \wp\ as in Figure \ref{xi}. Once scaled to this new value of
$\om$, the amplitude and scale dependence of the PVD from the HOD
models is generally consistent with the 2dFGRS data. There is some
discrepancy at $r_p=0.1$ \hmpc, but this data point suffers from fiber
collision effects and incompleteness due to blending of objects in the
input catalog. A quantitative comparison between the data and the
models is difficult because \cite{hawkins_etal:03} analyze a
flux-limited sample, and we have not replicated the details of their
modeling procedure. However, a flux-limited ``prediction'' would be a
linear combination of the different model curves (and the
cross-correlated velocity statistics). All the (scaled) model curves
show a smooth increase in $\sv(r_p)$ from $\sim 400$ \kms\ to $\sim
600$ \kms, in qualitative agreement with the data.

\section{Interpreting the Results of the Dispersion Model}

JB and L05 measure velocity statistics for volume-limited samples,
enabling easier comparison to theoretical predictions. However, the
quantity that they measure in not the PVD itself but the parameter
$\sk$ of the dispersion model, which convolves linear theory with an
exponential distribution of random galaxy velocities. With this model,
the redshift-space power spectrum $P_Z(k,\mu)$ is related to the
real-space power spectrum $P_R(k,\mu)$ by

\begin{equation}
\label{e.dispersion}
P_Z(k,\mu) = P_R(k)\left( 1 + \beta\mu^2\right)^2
\left( 1 + \sigma_k^2\mu^2k^2/2 \right)^{-1},
\end{equation}

\noindent where $\mu$ is the cosine of the angle of the wavevector
with the line of sight, $\beta = \om^{0.6}/b_g$, and $\sigma_k$ is the
non-linear galaxy velocity dispersion. The exponential
distribution assumed in equation (\ref{e.dispersion}) is both
observationally and theoretically motivated
(\citealt{davis_peebles:83, sheth:96}). In a recent paper,
\cite{roman:04} presents a thorough critique of the dispersion model,
pointing out several deficiencies. First, the physical probability
distribution of random velocities implied by equation
(\ref{e.dispersion}) is not a true exponential dispersion, but in fact
contains a discontinuity at the mode. Second, the velocity
distribution function of dark matter deviates significantly from an
exponential at some scales. This is also true of halo pairwise
velocities (\citealt{zurek_etal:94}). Inside the scale of the one-halo
term, an exponential is a good approximation for galaxies, but in the
two-halo term some deviations are to be expected. Third, equation
(\ref{e.dispersion}) is also only a true convolution if the parameter
$\sigma_k$ is assumed to be constant with scale, but this assumption
is contradicted by the results of Figure \ref{sigr}. The scale
dependence of the PVD is part what JB and L05 wish to probe,
thereby removing one of the theoretical underpinnings of equation
(\ref{e.dispersion}). In light of this, it is necessary to quantify
the accuracy of the JB implementation of the dispersion model in order
to assess the possible systematic errors accrued through its
use. \cite{jing_borner:01b} test the dispersion model with mock galaxy
distributions, finding favorable results, but they apply these tests to
multiple realizations of one distribution and not as a function of
luminosity. Here we test the dispersion model against all nine
magnitude bins.


\begin{figure*}
\centerline{\epsfxsize=5.5truein\epsffile{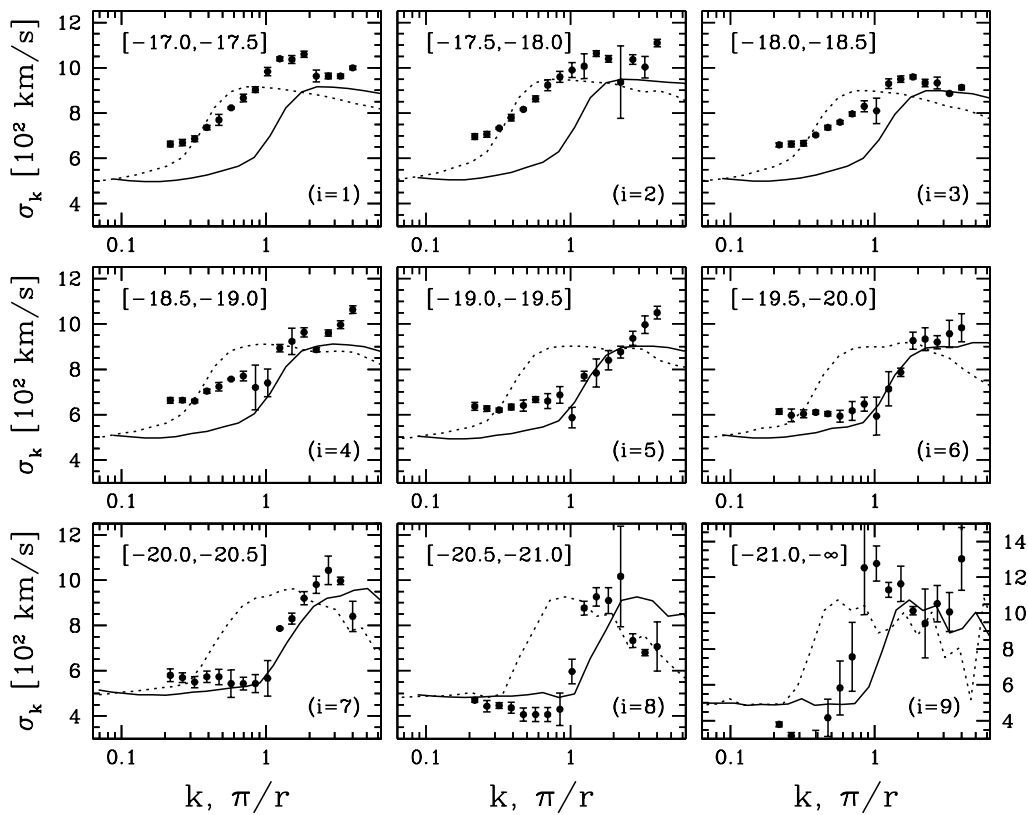}}
\vspace{-1.8cm}
\caption{ \label{sigk_sigr} Dispersion model results for the fiducial
  HOD model. Each panel represents a magnitude bin as labeled in
  top-left corner. Filled circles plot $\sk$ as a function of $k$, and
  solid lines represent the true PVD measured directly from the
  simulation (see Figure \ref{sigr}) plotted as a function of
  $\pi/r$. Error bars are the dispersion about the mean of the three
  projections of the simulation volume. The dotted lines plot the same
  results as the solid lines, but as a function of $1/r$. Note that
  the range on the $y$-axis has been increased to a maximum of 1,500
  \kms\ for the bottom right panel ($i=9$), which plots the results
  for the brightest magnitude bin.  }
\end{figure*}

We calculate the correlation function in redshift space $\xirp$, where
$r_p$ is projected separation and $r_\pi$ is line-of-sight
separation. We measure $\xirp$ for $0\le r_p\le 50$ \hmpc, and $-40
\le r_\pi \le +40$ \hmpc. We Fourier transform into $\pz$ by

\begin{equation}
\label{e.fft}
P_Z(k_p,k_\pi) = 2\pi \int dr_\pi \int dr_p \xirp r_p \cos(k_\pi r_\pi)J_0(k_pr_p)W_g(s)
\end{equation}

\noindent where $k_p$ and $k_\pi$ are the projected and line-of-sight directions in $k$-space, $J_0(x)$ is the zeroth-order Bessel function, and 

\begin{equation}
\label{e.weight}
W_g(s) = \exp\left(\frac{-s^2}{2S^2}\right),
\end{equation}

\noindent is a weighting function utilized to reduce the influence of
large scale distortions, where $s^2 = r_p^2 + r_\pi^2$, and $S=20$
\hmpc\ (see equations [6] and [8] in JB). We calculate the error in
$\pz$ from eight jackknife samples and fit equation
(\ref{e.dispersion}) at each $k$, determining the best-fit values of
$P_Z(k,0)$ and $\sk$ by $\chi^2$ minimization. We fix the value of
$\beta$ for each magnitude bin using the value of $b_g$ from Table 1,
but in practice we find the results are insensitive to the value of
$\beta$ (in agreement with JB and L05). 

Figure \ref{sigk_sigr} presents the results of the JB method as
applied to our fiducial model. The filled circles plot the dispersion
model results as a function of $k$, the solid lines plot the true PVD
in configuration space (from Figure \ref{sigr}) as a function of
$\pi/r$, and the dotted lines plot the same curves but as a function
of $1/r$. The errors on the solid points are the dispersion between
the three projections of the simulation volume. For all magnitudes,
the dispersion model results show the scale dependence expected from
Figure \ref{sigr}; at large scales (small $k$), there is an asymptotic
value of $\sk$, and at smaller scales the dispersion recovered from
the equation (\ref{e.dispersion}) rises, reaching a peak between
$k=1-3$ \mpch\ for most bins. For $i=4$ and $5$, $\sk$ has no clear
maximum value and continues to increase monotonically with
$k$. Qualitatively, the dispersion model correctly reproduces the
scale-dependence of the pairwise velocity dispersion that is measured
in configuration space. In detail, however, the comparison between
$\sk$ and $\sv$ is problematic. For the lowest three magnitude bins,
the transition from two-halo to one-halo pairs is complete by $k\sim
1$ \mpch, making a favorable comparison with the $\sv(1/r)$ (dotted)
curves, as shown in \cite{jing_borner:01b}. For brighter samples, this
transition moves to larger $k$, such that for $i=6-8$, $\sk$ clearly
follows the $\sv(\pi/r)$ (solid) curves. For the other samples, it is
not clear that either line best represents the dispersion model
results. The results of Figure \ref{sigk_sigr} underscore the
difficulty in translating the results of the dispersion model at a
given $k$ to a single physical scale. For faint galaxies, the
asymptotic value of $\sk$ at small $k$ is larger than the
high-separation value of $\sv$ by $\sim 200$ \kms, or $40\%$. For
samples near $\mstar$, $\sk$ and $\sv$ roughly agree at small $k$. For
the brightest samples, $\sk$ is smaller than $\sv$ at large scales.

\begin{figure*}
\centerline{\epsfxsize=4.5truein\epsffile{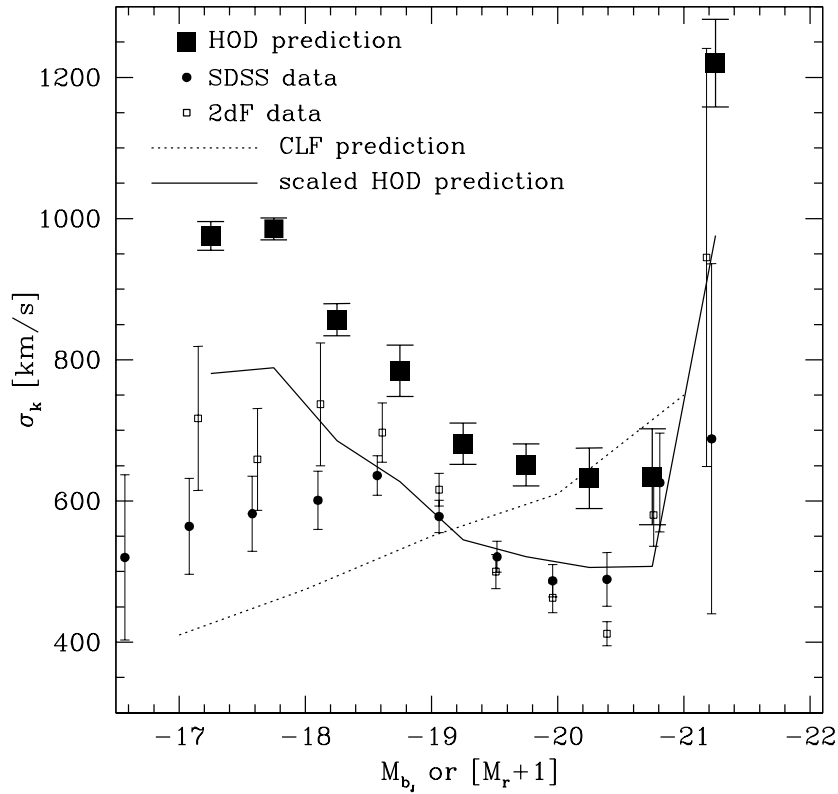}}
\caption{ \label{pvd_kspace} Comparison of the fiducial HOD model to
  the observational results of JB and L05. Open squares and filled
  circles show the measurements of JB and L05, respectively. The
  dotted line plots the CLF prediction shown in both papers. The large
  filled squares are the dispersion model results of the fiducial HOD
  model, averaged over the three $k$ values nearest $k=1$ \mpch. Error
  bars are the error in the mean. The solid line shows the HOD results
  shifted by a factor of 0.8 to represent a universe with $\om=0.2$.}
\end{figure*}

Figure \ref{pvd_kspace} compares the measurements of JB and L05 at
$k=1$ \mpch\ with the dispersion model results of the fiducial model
from Figure \ref{sigk_sigr}. The open squares and the solid circles
are the JB and L05 results, respectively. The results of both JB and
L05 show a clear trend of lower velocity dispersion for increasing
galaxy luminosity in the range $\mstar+1\ge \mbj \ge \mstar-1$. The
claim of JB is that this result is counterintuitive; brighter galaxies
should reside in higher-mass halos and therefore should be subject to
larger virial motions. This claim is reinforced by the predictions of
a conditional luminosity function model of the 2dFGRS by
\cite{yang_etal:03}. The dotted line in Figure \ref{pvd_kspace} plots
the predictions of their CLF model, obtained from inspection of Figure
9 in JB and Figure 7 in L05. The CLF prediction is nearly orthogonal
to the observational results, with a monotonic trend of increasing
$\sk$ with luminosity.

The large solid squares plot the HOD dispersion model results taken
from Figure \ref{sigk_sigr}. For several of the magnitude bins, the
value of $\sk$ at $k=1$ is slightly off the trend shown by the other
points, so we average the model results at $k=1$ with the data points
on either side. The error bars represent the error in the mean of the
nine values (three projections for each of three values of $k$). In
configuration space, there is little to no trend of $\sv$ with
luminosity at $r=1$ \hmpc\ (Figure \ref{pair_frac}c). The dispersion
model results of the fiducial model show a strong trend with
luminosity that is consistent with the dependencies seen in the JB and
L05 measurements. Figure \ref{sigk_sigr} demonstrates that at $k=1$
\mpch, $\sk$ for faint galaxies probes the regime where satellite
pairs dominate the statistic, while for brighter galaxies the $k=1$
\mpch\ scale is outside the one-halo regime, where the dispersion is
lower. This transition produces an enhanced luminosity dependence in
$\sk$ at this scale. The slope of the HOD results does not appreciably
change if we plot $\sk$ at $k=1$ \mpch\ without averaging over the
neighboring $k$ values.

As in Figure \ref{sigrp}, the HOD predictions for this cosmological
model are higher than the measurements. The solid line represents the
HOD model, scaled again by $(0.2/0.3)^{0.55}=0.8$. This scaled version
of our HOD prediction is consistent with the data in both slope and
amplitude, even reproducing the sharp upturn of the PVD for the
brightest galaxy sample. \cite{jing_borner:01b} applied the dispersion
model to the Las Campanas Redshift Survey, also finding that a model
with $\om=0.2$ was a better fit to the data than $\om=0.3$ (for
$\s8=1$).

The expectation of JB that brighter galaxies should occupy more
massive halos is borne out by our HOD modeling. But this creates a
strong correlation between the luminosity of {\it central} galaxies
and halo mass, while for satellite galaxies the lowest luminosities
dominate the distribution (by number) within a given halo. The PVD of
faint galaxies is high because of many of them are satellites in high
mass halos, and the PVD of luminous galaxies is lower than naively
expected because many of them are central galaxies of massive halos
that do not inherit the halo velocity dispersion. The results of JB
and L05 are thus naturally explainable in the halo occupation context
through the effect of $\fsat$, discussed in the previous section, and
by the systematic effects of using the dispersion model to estimate
the PVD.

Why are the predictions of two similar methods (e.g., the HOD and the
CLF) so clearly at odds? The real-space clustering results of the JB
implementation of the CLF overestimate the observed power spectrum for
$k\gtrsim 1$ \mpch\ by a factor of $3-5$ for galaxies brighter than
$\mbj = -19.5$ (see their Figure 6). This indicates that the number of
bright satellite galaxies is also overestimated, leading to more
one-halo pairs at these magnitudes than is observed, as well as a
higher satellite fraction at these magnitudes. A model that predicts
too many bright satellites would predict an increasing PVD with
luminosity, as \cite{slosar_etal:05} demonstrate. In the
\cite{yang_etal:03} CLF, galaxy clustering data are used only at the
scale length $r_0$ of the correlation function, well outside the
one-halo term.  Therefore, models that overpredict small-scale
clustering are not disfavored by high $\chi^2$ values in their
analysis. JB make a similar point about the best-fit parameters of the
\cite{yang_etal:03} model, noting that a model with a steeper
faint-end slope of the CLF could be more consistent with the PVD
measurements. The differences between our HOD results and the CLF come
as some surprise because of the good agreement between the two methods
in previous analyses (\citealt{tinker_etal:05a, vdb_etal:03}). These
differences are therefore specific to the statistic being modeled
here. The utilization of small-scale clustering for determining the
occupation functions makes our approach more robust for predicting
other statistics at small scales. For statistics that require
knowledge of the mean properties of the galaxies within halos, such as
the mass-to-light ratio or the overall PVD of galaxies, there should
be less bias in CLF results.


\section{Summary}

In this paper we use the halo occupation distribution framework to
make robust predictions of the pairwise velocity dispersion of
galaxies through modeling of the galaxy two-point correlation
function. Our technique is to constrain the parameters of the halo
occupation function $\navg$ by matching observations of the
luminosity-dependent projected correlation function \wp, then populate
the halos of an N-body simulation with mock galaxies drawn from the
inferred HOD. In our analysis, we make a distinction between central
galaxies, which we assume move with the center of mass of the host
halo, and satellite galaxies, which we assume have a velocity
dispersion within each host halo equal to that of the dark
matter. This distinction is crucial to our predictions because the
pairs that involve central galaxies have a lower dispersion, so the
fraction of satellites strongly influences both the luminosity and
scale dependence of the PVD in our predictions. At large separations,
the PVD is largely scale independent because halo velocities vary
weakly over the scales probed here. At $r\approx 2$ \hmpc, the PVD
rapidly increases as satellite-satellite pairs from massive halos
dominate. At $r< 1$ \hmpc, the PVD decreases with smaller
separation because central-satellite pairs become more common.

We have focused on the luminosity dependence of the PVD in light of
recent claims that the measured luminosity dependence is in conflict
with the expectations from halo occupation models. JB and L05 measure
a PVD that decreases with luminosity, which is inconsistent with the
expectation that brighter galaxies should have higher dispersions
because they occupy only high-mass halos. We have shown that this
expectation is incorrect; bright galaxies occupy the centers of
high-mass halos, so their dispersions are not enhanced by the large
virial motions within those halos. The dependence of $\sv$ on $\mbj$
is strongly influenced by the fraction of galaxies that are satellites
at each magnitude, a quantity that is $\sim 25\%$ for faint galaxies
and rapidly declines for $\mbj\le\mstar$. In the CLF prediction for
the PVD shown in JB and L05, the PVD monotonically increases with
luminosity. This prediction does include the distinction between
central and satellite galaxies, but it does not utilize \wp\ data at
small scales to constrain the occupation function, so it cannot
accurately constrain $\fsat$. While we find that observational
uncertainties in \wp\ and degeneracies among HOD parameters allow some
range in our predictions of the PVD, none of our 1,000 MCMC models
predict a monotonic increase in $\sv$ with luminosity.

L05 infer a significant dependence of PVD on color as well as
luminosity, with red galaxies having a higher dispersions. L05
conclude that the dependence of the PVD with luminosity results
from a population of faint red galaxies in clusters. The large
difference in the PVD of faint red and blue galaxies can also be
accounted for through halo occupation models. The HOD analysis of SDSS
correlation functions in \cite{zehavi_etal:05} includes analysis of
red and blue galaxies. The derived occupation functions predict a
higher satellite fraction for red galaxies and a larger number of
faint red galaxies in rich systems (see their Figure 23). These
occupation functions would produce a substantially higher PVD for red
galaxies. The conclusion of L05 is consistent with our interpretation
of these data through $\fsat$; faint satellite galaxies are primarily
red objects, while faint central galaxies are primarily blue galaxies
in low-mass halos.

The PVD is physically interesting quantity, both in terms of cosmology
and galaxy formation, but it cannot be measured directly with
precision. It can be inferred from the redshift-space correlation
function by model fitting, but the results are model dependent and may
be affected by the assumption of scale independence of the PVD. We
have included a qualitative comparison to the \cite{hawkins_etal:03}
configuration-space estimate of the PVD for a flux-limited sample of
2dFGRS galaxies, but we have focused most of our attention on JB and
L05's estimate for volume-limited samples, which use the dispersion
model to infer the PVD form the redshift-space power spectrum. We have
tested their implementation of this model with our fiducial HOD model,
demonstrating that the effective scale in configurations space probed
by a given $k$ scale in Fourier space changes with the luminosity of
the sample, complicating the comparison to predicted trends. At low
luminosities, $\sk$ at $k=1$ \mpch\ probes the regime where one-halo
pairs dominate the statistics, while for higher luminosities this scale
probes the the two-halo regime. Therefore, while the luminosity
dependence of the PVD in configuration space is flat at $r=1$ \hmpc,
$\sk$ at $k=1$ \mpch\ follows a strong trend with luminosity,
comparable to the dependence measured by JB and L05.

For our chosen cosmology, with $\om=0.3$ and $\s8=0.9$, the
normalization of the predicted PVD is too high, both in configuration
space and in Fourier space. This discrepancy can be resolved by
lowering $\om$ to 0.2, thus reducing halo velocities by
$(0.2/0.3)^{0.6}$ and virial dispersions by $(0.2/0.3)^{0.5}$. Our
predicted PVD would also be lower for a lower value of $\s8$. At large
scale, $\sv$ scales roughly linearly with $\s8$, while in the one-halo
regime $\sv\sim \s8^{0.5}$ for models constrained to match the same
\wp. It is difficult to use the PVD measurements on their own to
constrain cosmological parameters, however, because the normalization
of our predictions also depends on the velocity bias of both central
and satellite galaxies. A more general model would allow the satellite
galaxy dispersion to be $\sigma_{\rm sat} = \alpha_v\sigma_{dm}$, and
the central galaxy dispersion to be $\sigma_{\rm cen} =
\alpha_{vc}\sigma_{dm}$. In principle, both $\alpha_v$ and
$\alpha_{vc}$ could depend on halo mass and galaxy luminosity because
of dynamical effects on the orbits of subhalos that host satellite
galaxies. \cite{vdb_etal:05} argue that $\alpha_{vc}\approx 0.25$ for
2dFGRS groups, but in tests we find that significantly higher values
of $\alpha_{vc}$ are required to alter our predictions. At small
scales $\sv \sim \alpha_v$, while at large scales $\sv \sim
\av^{0.5}$. Thus the normalization of our prediction can be altered by
velocity bias. For the $\om=0.3$, $\s8=0.9$ model used here,
$\alpha\approx 0.65$ is required to match the large-scale amplitude of
the 2dFGRS measurements in Figure \ref{sigrp}, a bias that is stronger
than that predicted by numerical simulations.

The scale dependence of the PVD predicted by the HOD is in general
agreement with that of \cite{hawkins_etal:03}; both
the predictions and observations show a smooth transition between the
one-halo and two-halo regimes as a function of projected
separation. The $\sk$ curves in both JB and L05 show an increase in
$\sk$ at $k\sim 1$ \mpch, but the measurements lack the strong
two-halo to one-halo transition in our HOD models (Figure
\ref{sigk_sigr}). Changes in $\av$ and $\s8$ alter the PVD differently
at different scales. The difference in scale dependence may imply a
value of $\av<1$, or a higher $\s8$ than that assumed here, although a
higher value of $\s8$ would exacerbate the normalization problem in
Figures \ref{sigrp} and \ref{pvd_kspace}.

Recent papers by \cite{gao_etal:05} and \cite{harker_etal:05} have
shown that halo formation times correlate with the clustering strength
of halos (at fixed mass), implying that formation epoch correlates
with local environment. Papers by \cite{wechsler_etal:05} and
\cite{zhu_etal:06} demonstrate that the number of subhalos correlates
with halo formation epoch. The implication for the HOD is that $\navg$
could also correlate with environment, an effect not taken into
account in our modeling of \wp. If this correlation is strong, the
occupation functions derived in Figure \ref{hod} will be
systematically off their true values. In terms of pairwise velocity
statistics, however, these errors would not significantly change the
predictions. First, the effect of \cite{gao_etal:05} and
\cite{harker_etal:05} is strong for low-mass halos, where $\nsati \ll
1$. Any errors accrued would therefore be localized to the central
occupation function. Because the halo-halo velocity dispersion is
insensitive to halo mass for masses less than the non-linear mass
scale, these types of errors would not alter our predictions. Second,
assuming that the satellite occupation functions are systematically
off by as much as the difference in $\nsati$ between adjacent
magnitude bins, the change in $\sv$ per half magnitude is $\sim 5\%$
or less for most separations. Lastly, tests using a hydrodynamic
simulation presented in \cite{yoo_etal:05} demonstrate that any
environmental variation of the HOD has little impact on the galaxy
correlation function and is not likely to significantly alter the
derived occupation functions for luminosity-defined samples. The
recent numerical results do underscore the need to quantify the effect
that environmental correlations have on predictions of halo occupation
models, but we do not expect changes to the qualitative conclusions in
this paper.

Ultimately, there is more information about the PVD in the full,
two-dimensional redshift-space correlation function than in $\sv$
from the streaming model or $\sk$ from the dispersion
model. \cite{tinker_etal:05b} show that modeling redshift-space
distortions at small and large scales can separately constrain
$\alpha_v$, $\sigma_8$, and $\om$. Detailed observational comparisons,
imposing the constraint that $\s8$ and $\om$ must be the same for all
samples, could reveal luminosity and mass dependence of $\av$,
providing physical insight into the dynamical evolution of galaxies in
groups and clusters. We will apply these approaches to 2dFGRS and SDSS
observations in future work.

\vspace{1cm}

JT would like to thank Shaun Cole, Charlie Conroy, Carlos Frenk,
Andrey Kravtsov, Hiranya Peiris, Risa Wechsler, Renbin Yan, Andrew
Zentner, and Zheng Zheng for useful discussions. JT would also like to
thank the generous hospitality of the Institute for Computational
Cosmology at the University of Durham, where part of this work was
completed. PN acknowledges support of a PPARC PDRA fellowship and the
computer resources of ETH Zurich. DW acknowledges the support of NSF
grant AST-0407125. Portions of this work were performed under the
auspices of the U.S. Dept. of Energy, and supported by its contract
\#W-7405-ENG-36 to Los Alamos National Laboratory.  Computational
resources were provided by the LANL open supercomputing initiative.


\bibliography{../risa}


\end{document}